\documentclass[twocolumn,showpacs,prb,amsmath,amssymb,floatfix]{revtex4-1}
\usepackage{amsmath,amssymb,color}

\usepackage{bm}
\usepackage{epsfig}
\usepackage{psfrag}

\usepackage{verbatim}
\usepackage{lipsum}

\newcommand{\bd}{\bm}

\begin{document}

\title{Physical dipoles and
second order perturbation theory for dipolar fermions in two dimensions}

\author{Philipp Lange}
\affiliation{Institut f\"{u}r Theoretische Physik, Universit\"{a}t Frankfurt, Max-von-Laue Strasse 1, 60438 Frankfurt, Germany}

\author{Jan Krieg}
\affiliation{Institut f\"{u}r Theoretische Physik, Universit\"{a}t Frankfurt, Max-von-Laue Strasse 1, 60438 Frankfurt, Germany}

\author{Peter Kopietz}
\affiliation{Institut f\"{u}r Theoretische Physik, Universit\"{a}t Frankfurt, Max-von-Laue Strasse 1, 60438 Frankfurt, Germany}

 \date{\today}

\begin{abstract}
In two dimensions the Fourier transform  of the
interaction between two point dipoles has a term which grows linearly 
in the modulus $ | \bd{q}|$ of the momentum. As a consequence, in second order
perturbation theory
the self-energy  of two-dimensional dipolar fermions is ultraviolet divergent.
We show that for electric dipoles
this divergence can be avoided if one takes into account that
physical dipoles consist of
two opposite charges which are separated by a finite distance.
Using this regularization, we calculate the self-energy, the renormalized chemical potential,
and the renormalized Fermi surface of dipolar fermions in two dimensions in second order
perturbation theory.
We find that in the Fermi liquid phase the second order corrections
weaken first order effects.
\end{abstract}


\maketitle

\section{Introduction}

In the last decade many-body systems consisting of
fermionic atoms or molecules which carry a finite electric or magnetic dipole moment
have been intensely studied both 
experimentally~\cite{Sage05,Chi06,Ospelkaus08,Ni08,Ospelkaus09,Deiglmayr08,Deiglmayr10,Ni10,Miranda11,Heo12,Wu12,LuM12,Repp13,Aikawa14s,Aikawa14,Burdick15}
 and theoretically\cite{Bruun08,Baranov08,Cooper09,Fregoso09,Yam10,Zhao10,Liao10,Fregoso10,He11,Sieberer11,Liu11,Mikelsons11,Matveeva12,Gads12,Parish12,Zinner12,Volosniev13,Fregoso13,Bhongale13,Block14,Krieg15,Gras15,Bruun15,Wu15}.
In these systems the spatially anisotropic nature of the dipole-dipole
interaction gives rise to a rather complex phase diagram, which includes 
a normal Fermi liquid phase and phases with spontaneously
broken symmetries, such as  superfluid, magnetic, and nematic  phases.
Experimentally, many-body systems of fermions
coupled by dipolar forces
have been realized
both in two and three dimensions. 
Here we shall focus on many-body corrections to  
the single-particle spectrum in the normal Fermi liquid phase 
of dipolar fermions
in two spatial dimensions (2D).
Specifically, we shall calculate the 
irreducible self-energy $\Sigma ( \bd{k} , \omega )$ as a function
of momentum $\bd{k}$ and frequency $\omega$
up to second order  perturbation theory and determine the renormalized Fermi surface
to this order. Recently we have performed analogous calculations
for dipolar fermions in three dimensions \cite{Krieg15}. 
However, these calculations cannot be simply
extended to the two-dimensional case because in 2D the self-energy
 $\Sigma ( \bd{k} , \omega )$ is ultraviolet divergent.
This divergence is due to the fact that in momentum space the 2D  Fourier transform 
of the dipolar interaction
contains a term which grows linearly with the modulus of the momentum.
While several regularization strategies have been proposed in the literature~\cite{Fischer06,Chan10},
we show here that all divergences can be avoided if one 
replaces the point dipoles by physical dipoles.
In the electric case, we therefore replace each point dipole
by two opposite charges which are separated by a finite distance $\bd{\ell}$.
Using this regularization, we shall in this work calculate
the self-energy, the renormalized chemical potential, and the 
renormalized Fermi surface of two-dimensional dipolar fermions in second order perturbation
theory.

\section{Regularization of the dipolar interaction in 2D: physical dipoles}
\label{sec:regularizatrion}

\begin{figure}
\includegraphics[width=0.9\linewidth]{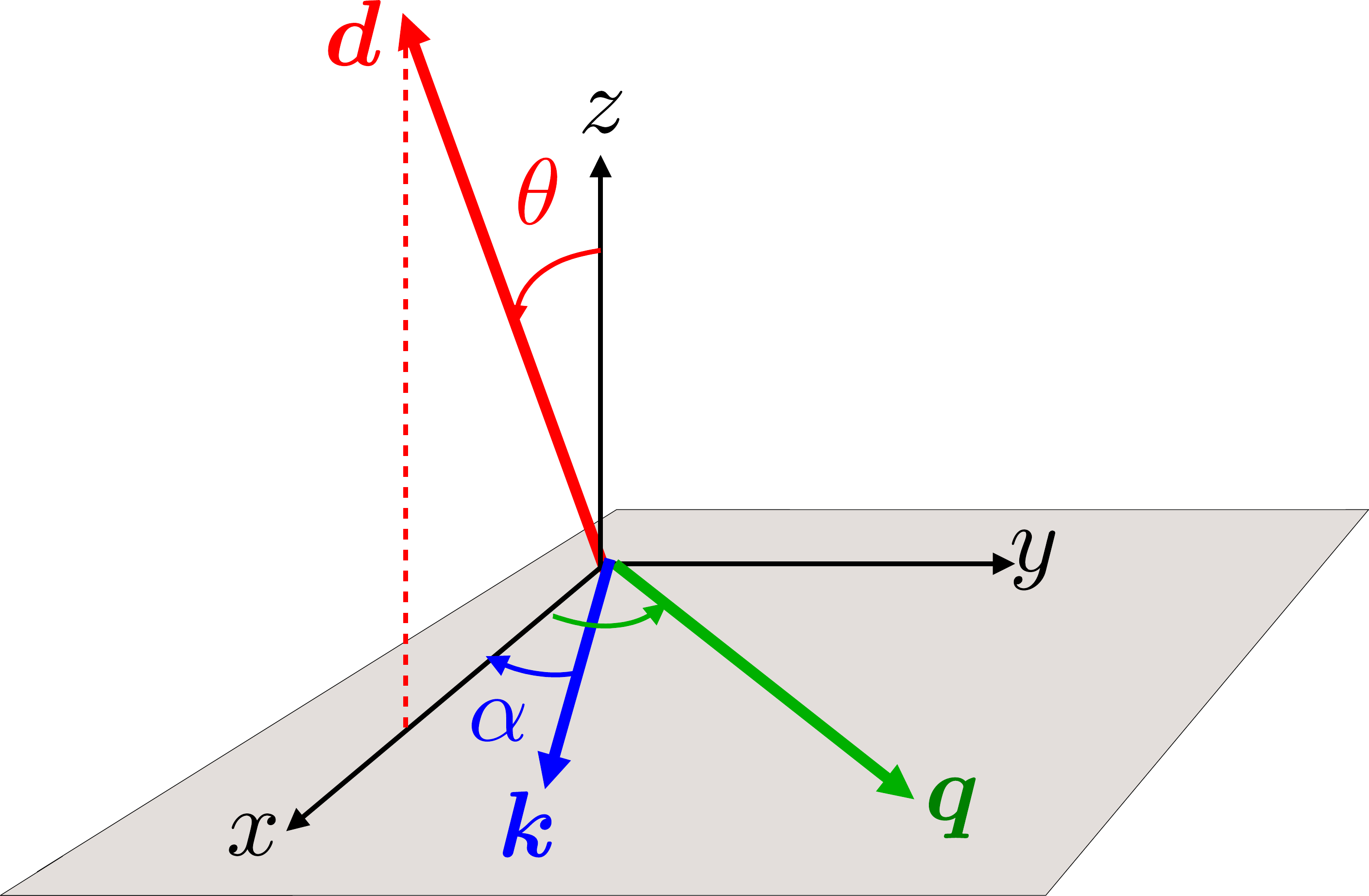}
\caption{
(Color online) 
For a two-dimensional system of dipoles $\bd{d}$ with arbitrary orientation, we choose our 
coordinate system such that the particles move in the $xy$-plane and $\bd{d}$ lies
in the $xz$-plane. The momentum transferred by the interaction is denoted
by $\bd{q}$. For later reference, we have also indicated the momentum
$\bd{k}$ of a quasi-particle. 
}
\label{fig:dipole2}
\end{figure}

Consider a many-body system consisting of spinless
fermionic atoms or molecules with mass $m$ and a finite dipole moment $\bd{d}$.
Assuming that the particles are confined to the $xy$-plane
and that all dipoles are aligned by an external field,
the second-quantized Hamiltonian of the system is
\begin{eqnarray}
 {\cal{H}} & = &   \int d^2 r \, \hat{\psi}^{\dagger} ( {\bd{r}} )
  \left( - \frac{\nabla^2}{2m} \right) \hat{\psi} ( {\bd{r}} ) 
 \nonumber
 \\*
 & + &
 \frac{1}{2} \int d^2 r \int d^2 r^{\prime} \,
\hat{\psi}^{\dagger} ( {\bd{r}} )\hat{\psi} ( {\bd{r}} )
 U ( {\bd{r}} - {\bd{r}}^{\prime} )  \hat{\psi}^{\dagger} ( {\bd{r}}^{\prime} )   
\hat{\psi} ( {\bd{r}}^{\prime} ),
 \nonumber
 \\*
 & &
 \label{eq:Hamiltonian}
\end{eqnarray}
where $\hat{\psi}^{\dagger} ( {\bd{r}} )$ and $\hat{\psi} ( {\bd{r}} )$ are the usual field 
operators which create or annihilate a fermion at position ${\bd{r}}
= r_x \hat{\bd{x}} + r_y \hat{\bd{y}}
$ in the $xy$-plane, and we have set $\hbar=1$. Here the orthogonal unit vectors $\hat{\bd{x}}$ and $\hat{\bd{y}}$ 
form a basis in the $xy$-plane.
The dipole-dipole interaction between two point dipoles with dipole moment $\bd{d}$ 
which are separated by the vector $\bd{r}$ is given by
 \begin{align}
U (\bd{r}) =\frac{d^{2}}{|\bm{r}|^{3}} \left[ 
 1 - 3 ( \hat{\bd{d}} \cdot \hat{\bd{r}} )^2 \right],
 \label{eq:rspointdipole}
 \end{align}
where $\hat{\bd{d}} = \bd{d} / | \bd{d} |$ and
 $\hat{\bd{r}} = \bd{r} / | \bd{r} |$ are unit vectors.
Expanding the field operators in momentum space,
$\hat{\psi} ( \bd{r} ) = \frac{1}{\sqrt{V}} \sum_{\bd{k}} e^{ i \bd{k} \cdot \bd{r} } 
 c_{\bd{k}}$, where $V$ is the (two-dimensional) volume of the system,
our Hamiltonian (\ref{eq:Hamiltonian}) can be written as
 \begin{eqnarray}
 {\cal{H}} & = & \sum_{\bd{k}} \frac{ \bd{k}^2}{2m} c^{\dagger}_{\bd{k}} c_{\bd{k}}
+ \frac{1}{2V} \sum_{\bd{k}, \bd{k}^{\prime} \bd{q}} 
 U_{\bd{q}} c^{\dagger}_{ \bd{k} + \bd{q}} c^{\dagger}_{\bd{k}^{\prime} - \bd{q}}
 c_{\bd{k}^{\prime} } c_{\bd{k}},
 \end{eqnarray}
where the two-dimensional Fourier transform of the dipolar interaction is defined by
 \begin{equation}
 U_{\bd{q}} = \int d^2 r e^{ -i \bd{q} \cdot \bd{r} } U ( \bd{r} ).
 \label{eq:Uqdef}
\end{equation}
As it stands, Eq.~(\ref{eq:Uqdef}) is ultraviolet divergent and has to be regularized.
One possibility is to exclude the short-distance regime $| \bd{r} | < r_0$
from the $\bd{r}$-integration. 
This regularization has been introduced in Ref.~[\onlinecite{Chan10}].
The integral is then finite and one obtains
the regularized Fourier transform of the dipolar interaction in 2D
for small momenta $ | \bd{q} | \ll 1/ r_0$,
 \begin{eqnarray}
 U_{\bd{q}}^{(r_0)} & = &  \int_{| \bd{r} | > r_0}  d^2 r e^{ -i \bd{q} \cdot \bd{r} } 
 U ( \bd{r} )
 \nonumber
 \\
 &\approx  & - \frac{\pi d^2}{r_0} \left[ 1 - 3 \cos^2 \theta \right]
 \nonumber
 \\
 & & 
 - 2 \pi d^2 | \bd{q} | \left[ \cos^2 \theta - ( \hat{\bd{x}} \cdot \hat{\bd{q}} )^2
 \sin^2 \theta \right],
 \label{eq:FT1}
 \end{eqnarray}
where $\theta$ is the angle between $\bd{d}$ and the $z$-axis and we have oriented our
coordinate system in the $xy$-plane such that $\bd{d}$ lies in the $xz$-plane,
as shown in Fig.~\ref{fig:dipole2}.
Although the first term in Eq.~(\ref{eq:FT1})  explicitly depends
on the cutoff $r_0$, this term corresponds to a contact interaction
and does not affect the physics of a one-component Fermi gas.
Keeping in mind that the linear increase of the second term
in Eq.~(\ref{eq:FT1}) should be cut off at $ | \bd{q} | \approx 1/ r_0$,
perturbation theory with 
the interaction 
 $U_{\bd{q}}^{(r_0)}$ is ultraviolet convergent; however, 
physical quantities which depend on short-distance physics 
can exhibit a rather complicated dependence on $r_0$.

The above complications
can be avoided if one uses the
regularization procedure proposed by Fischer~\cite{Fischer06}
who proposed to consider a quasi-two-dimensional system where the dipoles
are confined to the lowest transverse subband
due to an external harmonic confining potential in $z$-direction. 
If the trapping frequency $\omega_z$ of the confining potential is larger than the chemical potential $\mu$ and the temperature $T$, the particles 
occupy only the lowest transverse subband  with wavefunction
 \begin{equation}
\phi_0 (z)=\pi^{-1/4} a^{-1/2}e^{-z^2/(2a^2)}.
 \end{equation}
The length scale $a$ is defined by the trapping frequency $\omega_z$ 
and is given by $a=(m \omega_z )^{-1/2}$. 
One then convolutes the three-dimensional dipolar interaction $ U ( \bd{r} - \bd{r}^{\prime} , z - z^{\prime} )$ with the  probability density profile in the transverse direction,
 \begin{align}
  U^{(a)}(\bd{r} )= \int dz \int dz^{\prime}\phi_0 (z)^2\phi_0 (z^{\prime})^2 
 U(\bd{r},z-z^{\prime}).
\label{eq:rsquasi2D}
 \end{align}
The integration can be carried out exactly;
Fourier transforming the result to momentum space one obtains
for the regularized effective 2D dipolar interaction
 \begin{align}
  U^{(a)}_{\bd{q}} & = 
  - \frac{\sqrt{2\pi} d^{2}}{a}
 \left[ 1 - 3 \cos^2 \theta \right]
  \nonumber
  \\
  & -\frac{2\pi d^{2}}{a}   F ( | \bd{q} | a )  
 \left[\cos^2 \theta - ( \hat{\bd{x}} \cdot \hat{\bd{q}} )^2
 \sin^2 \theta \right],
 \label{eq:ftquasi2D}
 \end{align}
where the dimensionless function $ F (x)$ can be expressed 
in terms of the complementary error function Erfc$(x)$ as follows,
\begin{align}
 F(x)=x e^{x^2 / 2}\text{Erfc}(x/ \sqrt{2}).
\end{align}
Note that $F ( x ) \sim x$ for small $x$ so that
in the long-wavelength regime
$ | {\bd{q}}  |  \ll 1/a$  the momentum-dependent part of 
$ U^{(a)}_{\bd{q}}$ agrees precisely with the
momentum-dependent part of $U_{\bd{q} }^{(r_0)}$
in Eq.~(\ref{eq:FT1}).
On the other hand, using 
the fact that $F ( x ) \sim \sqrt{2 / \pi } + {\cal{O}} ( x^{-2} )$ for large $x$,
we see that for $ | \bd{q} |  \gg 1/a$ the regularized interaction
  $U^{(a)}_{\bd{q}}$ approaches a finite limit,
\begin{align}
\lim\limits_{|\bm{q}|\rightarrow \infty}
U^{(a)}_{\bd{q}} = \frac{\sqrt{2\pi}d^2}{a}\left[2(\hat{\bd{x}} \cdot \hat{\bd{q}})^2-1\right]\sin^2\theta ,
\end{align}
which is independent of $ | \bd{q} |$ but
still depends on the direction $\hat{\bd{q}}$.
The above regularization of the dipolar interaction in two dimensions  has recently
 been used by several
authors~\cite{Bruun08,Tic11,Sieberer11,Babadi12,Babadi13,Block14}.

\begin{figure}
\includegraphics[width=0.9\linewidth]{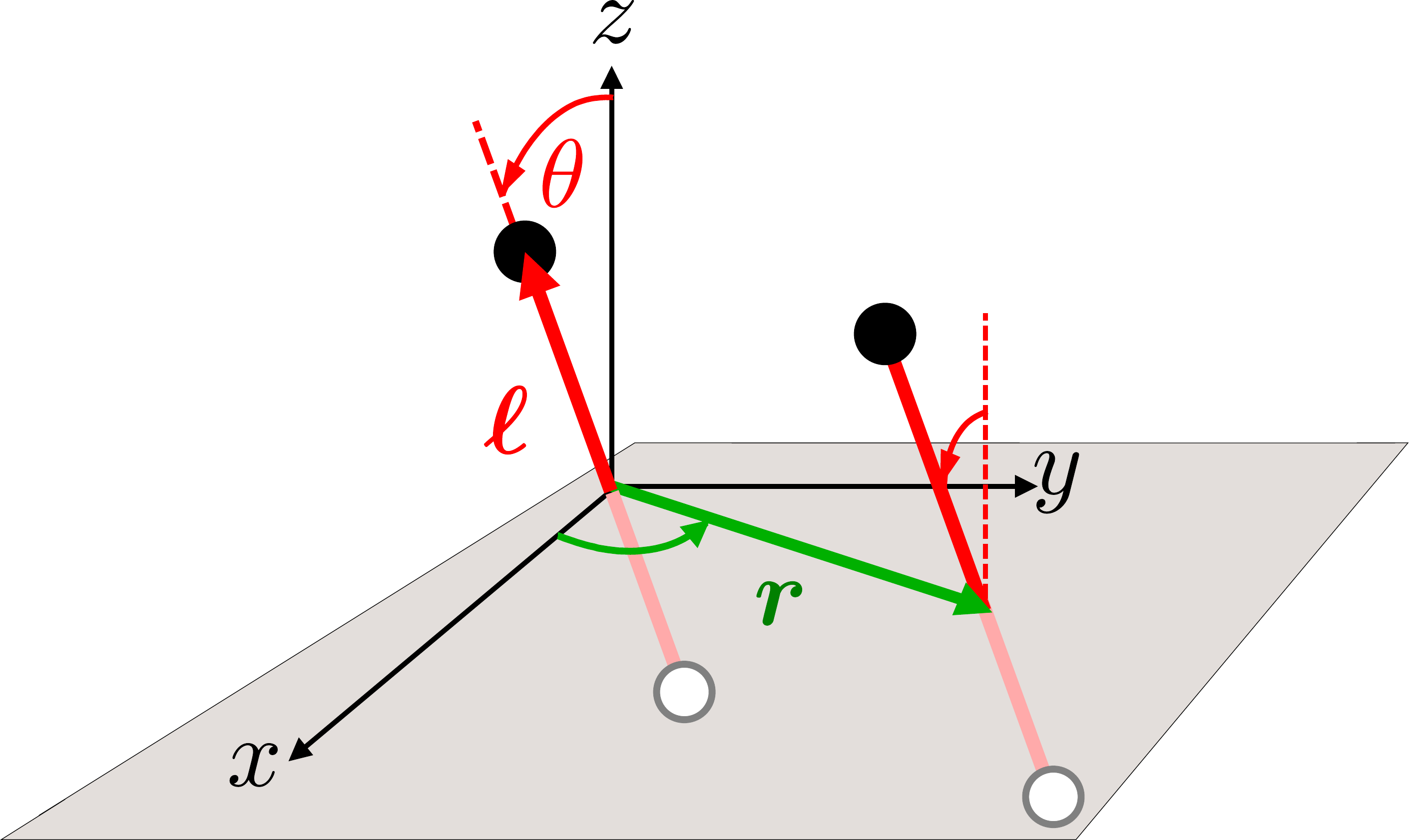}
\caption{
(Color online)
Sketch of two aligned dipoles in real space.
The displacement vector $\bd{\ell}$ lies in the $xz$-plane 
and the vector $\bd{r}$, separating the two dipoles, lies in the $xy$-plane.
The black and white dots represent the charges of the dipole. 
}
\label{fig:realspace_dipole}
\end{figure}

Let us now propose a different regularization procedure which does not require
the embedding of the system into a harmonic trapping potential and has the advantage
that the Fourier transform of the regularized interaction vanishes for large
momentum transfers. For simplicity, we focus on electric dipoles. 
A generalization
to magnetic dipoles seems to be  possible, although we have not attempted to
work out the details.
The crucial point is that a physical  electric dipole consists of two 
opposite point charges $\pm Q$ which are separated by a finite distance $\bd{\ell}$.
\footnote{Instead of representing a physical dipole by point charges, one could alternatively use a continuous smeared out charge density 
to regularize the interaction, see B. P. van Zyl, E. Zaremba, and P. Pisarski, Phys. Rev. A \textbf{87}, 043614 (2013).}
The corresponding electric dipole moment is $\bd{d} = Q \bd{\ell}$.
The three-dimensional charge density at position $\bd{R}$
of $N$ aligned electric dipoles whose center is located at
positions $\bd{R}_i$ is therefore given by
 \begin{eqnarray}
 \rho^{(3)} ( \bd{R} ) & = &  Q \sum_{i=1}^N 
 \bigl[ \delta^{(3)}  ( \bd{R} - \bd{R}_i - \bd{\ell} /2  )
 \nonumber
 \\
 & & \hspace{7mm}
 - \delta^{(3)} ( \bd{R} - \bd{R}_i +  \bd{\ell} /2 ) \bigr],
 \end{eqnarray}
where $\delta^{(3)} ( \bd{R} )$ is the three-dimensional Dirac $\delta$-distribution.
The electrostatic Coulomb energy stored in this charge distribution is
 \begin{equation}
 E_{\rm el} = \frac{1}{2} \int d^3 R \int d^3 R^{\prime} \frac{ \rho^{(3)} ( \bd{R} )
 \rho^{(3)} ( \bd{R}^{\prime} )}{ | \bd{R} - \bd{R}^{\prime} | }.
 \label{eq:Eel}
 \end{equation}
Suppose now that the centers of all dipoles are located in the $xy$-plane 
and that all dipole moments are aligned and lie in the
$xz$-plane, as shown in Fig.~\ref{fig:realspace_dipole}.
The positions of the centers of the dipoles are then of the form
$\bd{R}_i = \bd{r}_i$, where the vector $\bd{r}_i$ has no components 
in $z$-direction. Setting $\bd{R} = \bd{r} + z \hat{\bd{z}}$ and
  $\bd{R}^{\prime} = \bd{r}^{\prime} + z^{\prime} \hat{\bd{z}}$
in Eq.~(\ref{eq:Eel}),
where $\bd{r} $ and $\bd{r}^{\prime}$ lie now in the $xy$-plane,
and integrating over $z$ and $z^{\prime}$ we can 
rewrite the electrostatic interaction energy of the dipoles 
in the form
 \begin{eqnarray}
 E_{\rm el} & = & \frac{Q^2}{2} \sum_{ij} 
 \left[ \frac{2}{| \bd{r}_i - \bd{r}_j | }
 - \frac{1}{ | \bd{r}_i - \bd{r}_j + \bd{\ell} |}
 - \frac{1}{ | \bd{r}_i - \bd{r}_j - \bd{\ell} |} \right]
 \nonumber
 \\
 & = & 
\frac{1}{2} \int d^2 r \int d^2 r^{\prime} 
 \rho ( \bd{r} )
U^{(\ell )} (  \bd{r} - \bd{r}^{\prime} ) 
 \rho ( \bd{r}^{\prime} ),
 \end{eqnarray}
where
 \begin{equation}
 \rho ( \bd{r} ) = \sum_{i=1}^{N} \delta^{(2)} ( \bd{r} - \bd{r}_i )
 \end{equation}
is the two-dimensional density of dipoles and 
 \begin{equation}
 U^{(\ell )} (  \bd{r} ) = Q^2 \left[ \frac{ 2}{ | \bd{r} |}  -
 \frac{1}{ | \bd{r} + \bd{\ell} | } -  \frac{1}{ | \bd{r} - \bd{\ell} | }
 \right]
 \label{eq:Uell_real}
 \end{equation}
is the regularized dipole-dipole interaction.
In the isotropic case ($\theta = 0$) this reduces to the interaction 
used in Ref.~[\onlinecite{Fregoso13}]
to study Wigner crystallization in two dimensions.
The two-dimensional Fourier transform 
of Eq.~\eqref{eq:Uell_real} is
 \begin{equation}
 U_{\bd{q}}^{ ( \ell )} = \frac{ 4 \pi Q^2}{ | \bd{q} | }
 \left[ 1 - e^{ - \ell | \bd{q} | \left| \cos \theta \right| }
 \cos ( \bd{q} \cdot \hat{\bd{x}} \ \ell \sin \theta ) \right],
 \label{eq:Uell}
 \end{equation}
where $\theta$ is again the angle between the direction of the dipole and the $z$-axis,
see Fig.~\ref{fig:realspace_dipole}.
Keeping in mind that $Q = d / \ell$ and expanding the term in the square brackets of
Eq.~(\ref{eq:Uell}) up to second order in $|\bm{q}|$
we obtain
 \begin{eqnarray}
 U_{\bd{q}}^{ ( \ell )} & = & \frac{ 4 \pi d^2}{ \ell } \left|\cos \theta \right|
 \nonumber
 \\
 &  - & 2 \pi d^2 | \bd{q} | \left[ \cos^2 \theta - ( \hat{\bd{x}} \cdot \hat{\bd{q}} )^2
 \sin^2 \theta \right]
 + {\cal{O}} ( \bd{q}^2 ).
 \end{eqnarray}
\begin{figure}
\includegraphics[width=1\linewidth]{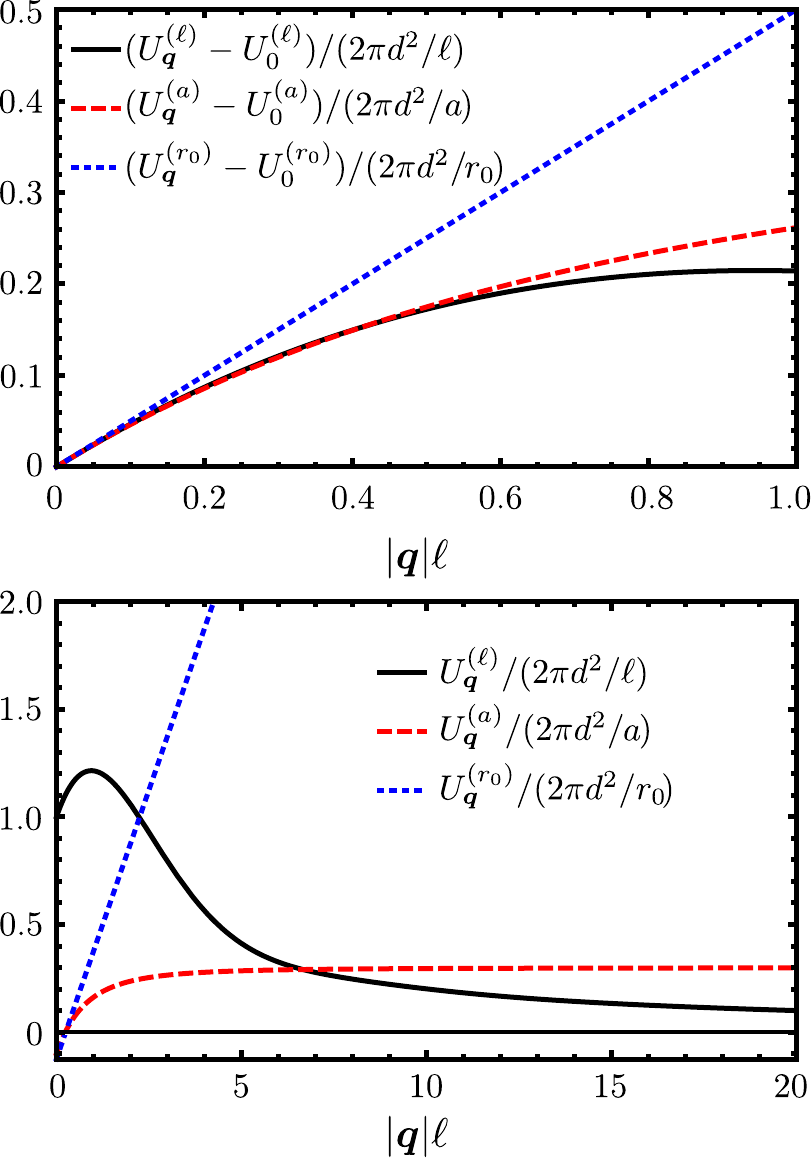}
\caption{
(Color online)
Plot of the Fourier transform of the dipole-dipole interaction
for tilt angle $\theta=\pi/3$ and $\hat{\bd{q}} \cdot \hat{\bm{x}}=1$ against $|\bm{q}| \ell$ for all three regularization strategies
discussed in Sec.~\ref{sec:regularizatrion};
$U_{\bd{q}}^{(\ell)}$ (black solid line), $U_{\bd{q}}^{(a)}$ (red dashed line), and $U_{\bd{q}}^{(r_0)}$ (blue dotted line). 
To compare the regularization strategies we set $r_0 = a = \ell$.
The upper panel shows the behavior for small $|\bm{q}| \ell$ where we have shifted $U_{\bd{q}}$ by $U_{0}$ for better comparison.
In the lower panel we plot $U_{\bm{q}}$ in a larger momentum range. 
}
\label{fig:compare}
\end{figure}
Obviously,  the leading order momentum-dependent term
agrees with the momentum-dependent term
of $U^{(r_0)}_{\bd{q}}$ in Eq.~(\ref{eq:FT1}).
On the other hand, for $| \bd{q} | \gg 1/ \ell $
the regularized Fourier transform (\ref{eq:Uell}) 
vanishes as $4 \pi Q^2 / | \bd{q} |$, which is just the bare Coulomb interaction
between the charges attached to two dipoles.
The $1/ | \bd{q} |$ decay of the interaction for large momenta guarantees that
perturbation theory in powers of $U^{(\ell)}_{\bd{q}}$ is ultraviolet convergent.
In Fig.~\ref{fig:compare} we compare the momentum dependence of the
three different regularized dipolar interactions given 
in Eqs.~(\ref{eq:FT1}), (\ref{eq:ftquasi2D}) and (\ref{eq:Uell}).
Once we subtract the regularization dependent constants $U_0^{(\ell)}$, $U_0^{(a)}$, or $U_0^{(r_0)}$, the behavior
for $\left| q \right| \ll1/\ell$ is very similar for all three regularization strategies.
Since we consider a one-component Fermi gas the self-energy does not depend on these constants.
However, for large momenta $U_{\bd{q}}^{(\ell)}$ vanishes as $|\bm{q}|^{-1}$,
whereas $U_{\bd{q}}^{(r_0)}$ grows linearly, and $U_{\bd{q}}^{(a)}$ approaches a finite limit which depends on 
$\theta$ and $\hat{\bm{q}}$.

\section{Renormalized Fermi surface and chemical potential}
\label{sec:results}

In this section we will work with the regularization via physical dipoles $U^{(\ell)}_{\bd{q}}$. For convenience we rename
$U^{(\ell)}_{\bd{q}} \rightarrow U_{\bd{q}}$,
i.e., we omit the superscript indicating the regularization dependence
of the interaction.
In the normal Fermi liquid phase, the self-energy
of dipolar fermions up to second order in the interaction is
in the thermodynamic limit and at vanishing temperature given by\cite{Krieg15}
 \begin{equation}
 \Sigma ( \bd{k} , i \omega  ) 
 = \Sigma_{1} ( \bd{k} ) + \Sigma_{2} ( \bd{k} , i \omega  ),
 \end{equation}
where the first order correction is
 \begin{align}
\Sigma_{1} ( \bd{k} ) = \int_{\bd{q}} 
 ( U_0 - U_{ \bd{q}} ) \Theta ( -\xi_{\bd{k} + \bd{q}} ),
 \label{eq:sigma1_limit}
 \end{align}
and the second order term is
 \begin{widetext}
 \begin{align}
 \nonumber
 \Sigma_{2} ( \bd{k} , i \omega  ) & =  
  -\int_{\bd{q}}
 \int_{ \bd{q}^{\prime}}
 (U_{0}-U_{\bd{q}})(U_{0}-U_{\bd{q}^{\prime}})\Theta (-\xi_{\bd{q}+\bd{q}^{\prime}}) \delta(\xi_{\bd{q}^{\prime}+\bd{k}})
 \\
 &+
 \frac{1}{2}  \int_{\bd{q}} \int_{\bd{q}^{\prime}}
  [    U_{\bd{q}}  -  U_{\bd{q}^{\prime} }  ]^{2}
 \frac{  \Theta ( \xi_{ \bd{k} + \bd{q}} )  \Theta ( \xi_{ \bd{k} + \bd{q}^{\prime}} )
  \Theta ( - \xi_{ \bd{k} + \bd{q} + \bd{q}^{\prime}} )  +
\Theta ( - \xi_{ \bd{k} + \bd{q}} )  \Theta ( - \xi_{ \bd{k} + \bd{q}^{\prime}} )
  \Theta (  \xi_{ \bd{k} + \bd{q} + \bd{q}^{\prime}} )
  }{i\omega  -  ( \xi_{ \bd{k} + \bd{q}} +  \xi_{ \bd{k} + \bd{q}^{\prime}} 
  - \xi_{ \bd{k} + \bd{q} + \bd{q}^{\prime}}         )     }.
 \label{eq:sigma2_limit}
 \end{align}
\end{widetext}
Here $\int_{\bd{q}} = \int d^2 q /(2 \pi)^2$ and
$\xi_{\bd{k}}=\bd{k}^2/(2m)-\mu$
is the bare energy  relative to the chemical potential $\mu$.
 The first term in Eq.~\eqref{eq:sigma2_limit} arises from the iteration
of the first order self-energy, while the frequency-dependent term in the second line
corresponds to the second order contributions which cannot be generated
by iterating first order corrections.

We have calculated the above self-energies numerically
using the VEGAS Monte Carlo algorithm 
from the GNU Scientific Library~\cite{GSL} for the numerical integrations in C$++$.
Given the self-energy $ \Sigma ( \bd{k}_F , 0 )$ at vanishing frequency,
we can determine the renormalized Fermi surface 
by solving the implicit equation
\begin{align}
 \frac{ k_F^2}{2m} + \Sigma(\bd{k}_F, 0)=\mu.
 \label{eq:selfconFS}
\end{align}
It is convenient to parametrize the
renormalized Fermi surface in terms of polar coordinates,
$ \bd{k}_F = k_F ( \alpha ) ( \cos \alpha \hat{\bd{x}} + \sin \alpha \hat{\bd{y}} )$,
where $\alpha$ is the angle between $\bd{k}_F$ and the $x$-axis, 
see Fig.~\ref{fig:dipole2}.
As usual, we work at constant density $n = k_{F0}^2 / ( 4 \pi )$,
where $k_{F0}$ is the Fermi momentum in the absence of interactions.
The chemical potential is then renormalized by  the interaction 
and it is convenient to introduce the interaction-dependent 
renormalization factor\cite{Krieg15}
\begin{align}
 \gamma^2 = \frac{\mu}{E_{F0}}=\frac{2 m \mu}{k_{F0}^2}.
\end{align}
The self-consistency equation (\ref{eq:selfconFS}) for the Fermi surface can
then be written as 
\begin{align}
\label{eq:kfdef}
\frac{k_{F}(\alpha)}{\gamma k_{F0}} =\sqrt{1-\frac{\Sigma( {k}_{F}(\alpha),0)}{\mu}},
\end{align}
where $\Sigma ({k}_F ( \alpha)  , 0 ) $ is an abbreviation for
$\Sigma ( k_F ( \alpha )  [  \hat{\bd{x}} \cos \alpha   +  \hat{\bd{y}} \sin \alpha   ] , 0 )$.
A dimensionless measure for the strength of the interaction
is 
\begin{align}
 u =  2 \pi \nu  d^2 k_{F0} = m  d^2 k_{F0},
 \end{align}
where $\nu = m /(2 \pi )$ is the density of states in two dimensions.
It is also convenient to introduce the dimensionless ultraviolet cutoff
$\Lambda = 1 / ( k_{F0} \ell )$.
Using the parameters realized in the experiment by
de Miranda \textit{et al.}~\cite{Miranda11}
for dipolar $^{40}$K$^{87}$Rb
in a 2D pancake geometry 
we work with an approximate density
of $n\sim 10^7 \text{cm}^{-2}$.
Keeping in mind that the size of dipolar molecules is of the order of $\ell \sim 10 \text{\AA}$
we estimate $\Lambda \approx 1000$ which we use for our numerical calculations.

\begin{figure}
\includegraphics[width=\linewidth]{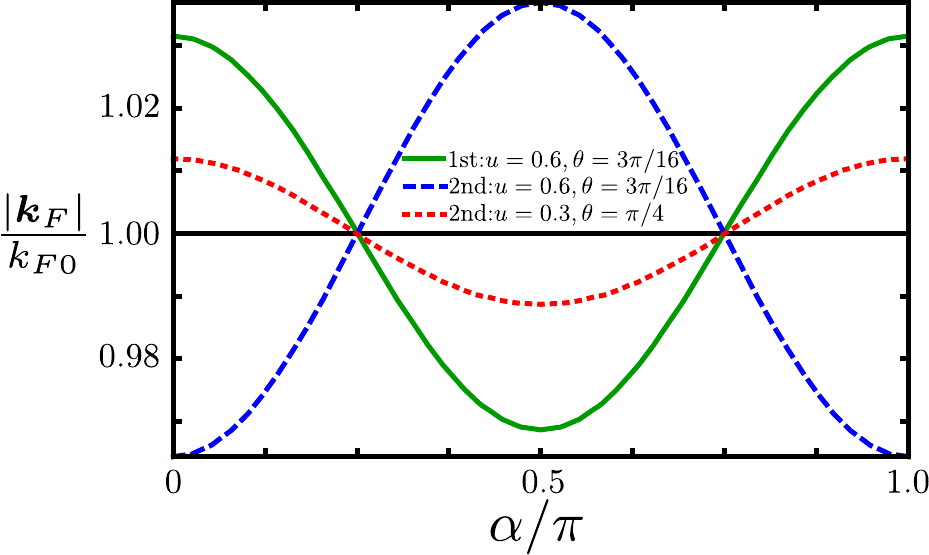}
\caption{
(Color online) Renormalized Fermi surface in the interval $0\le \alpha \le \pi$
to first and second order in perturbation theory for various choices of $u$ and $\theta$ 
and ultraviolet cutoff $\Lambda = 1/(k_{F0} \ell) = 1000$.
The solid green line represents the 
Fermi surface up to first and the dashed blue line up to second order in the interaction at $u=0.6$ and $\theta=3\pi/16$. 
The dotted red line is the Fermi surface up to second order in the interaction with $u=0.3$ and $\theta=\pi/4$. 
The undisturbed Fermi surface is plotted as a reference (solid black line).
As we can see, the Fermi surface shows a qualitatively different behavior for certain $u$ and $\theta$ in second order perturbation theory. 
Note that the tilt angle of $\theta=\pi/4$ is above the critical tilt angle for an instability towards the superfluid phase~\cite{Sieberer11,Parish12};
we have choosen this rather large tilt angle for a better visibility of the second order effect at $u=0.3$.
}
\label{fig:FS}
\end{figure}

Our results for the renormalized Fermi surface are shown in
Fig.~\ref{fig:FS}.
For sufficiently small interactions 
the distortion of the Fermi surface can be understood within first order 
perturbation theory.
To this order the  renormalized Fermi surface of dipolar 
fermions in 2D has already been 
calculated by Chan {\it{et al.}}~\cite{Chan10}. The Fermi surface
stretches along the direction given by the projection of the dipole onto the $xy$-plane,
which is the $x$-axis with our choice of the coordinate system.
Intuitively, this is due to the fact that for finite $\theta$ the interaction in real space
[Eq.~\eqref{eq:Uell_real}] is most attractive (or, depending on $\theta$, least repulsive)  
along the $x$-axis.
Therefore particles moving along the $x$-axis need less energy than particles moving transverse 
to this axis -- hence, the Fermi surface is stretched along the direction 
defined by the projection of the dipoles on the $xy$-plane.

\begin{figure}
\includegraphics[width=1\linewidth]{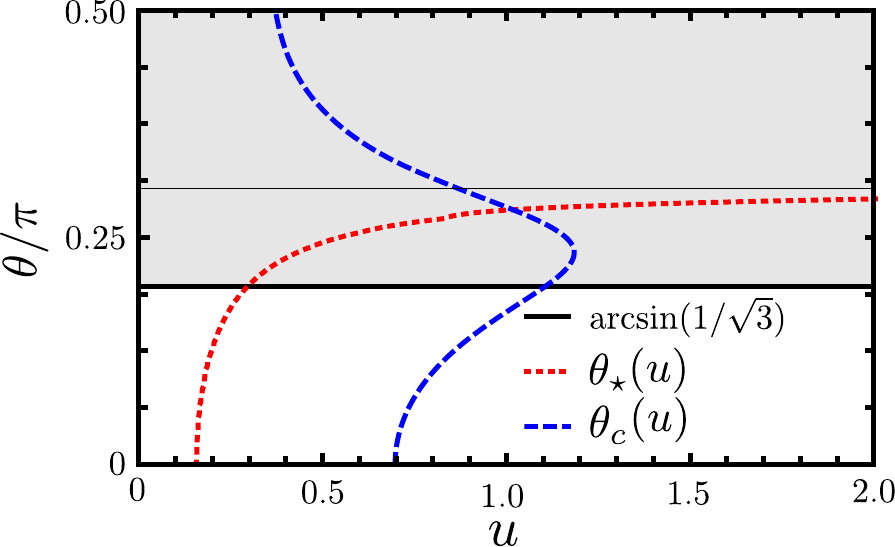}
\caption{
(Color online) 
Plot of $\theta_\star(u)$ (red dotted line) and $\theta_c(u)$ (blue dashed line),
the critical tilt angle for a vanishing bulk modulus,
in full second order perturbation theory. 
On the left hand side of $\theta_\star(u)$ the Fermi surface is stretched along the $x$-axes, on the right hand side along the $y$-axes. 
As we can see, the Fermi surface changes its elongation before the instability in the density-density channel occurs.
The lower horizontal line at $\theta=\arcsin {(1/\sqrt{3})}$ is approximately the critical tilt angle for an instability towards a superfluid 
phase~\cite{Bruun08,Parish12}.
}
\label{fig:beta_1_of_u}
\end{figure}

The second order correction to the self-energy tends  to 
weaken the first order distortion of the Fermi surface and has the tendency to distort the
Fermi surface in a direction that is perpendicular to the projection of the
dipoles onto the $xy$-plane, which is the $y$-direction with our choice of
coordinate system.
In fact, if we extrapolate the second order correction 
to intermediate values of the interaction, the second order correction
dominates the first order contribution so that the overall distortion of the
Fermi surface is along the $y$-direction.
If we choose the interaction parameter  $u$ to be larger than
$u_\star \approx 0.15$, we can always find a specific tilt angle
$\theta_\star ( u )$ of the dipoles such that for $\theta < \theta_\star ( u )$ the
Fermi surface is distorted along the $y$-axis.
The angle $\theta_\star ( u )$ can be defined
from the condition that $k_F ( \alpha =0) = k_F ( \alpha = \pi/2 )$ and
is shown as a dotted line in Fig.~\ref{fig:beta_1_of_u}.
It would be interesting to investigate whether this effect survives if 
the interaction is treated non-perturbatively.

\begin{figure}
\includegraphics[width=1\linewidth]{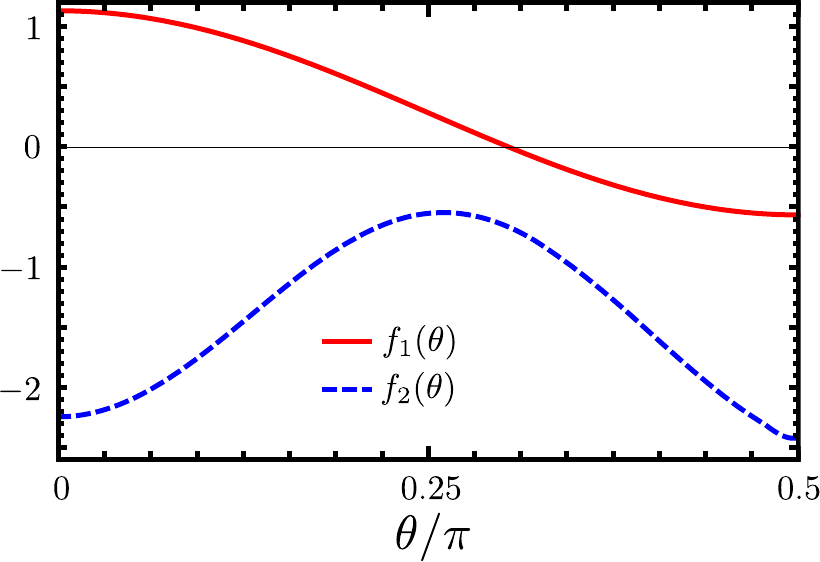}
\caption{
(Color online) Plot of the functions $f_{1}(\theta)$ and $f_{2}(\theta)$ from the perturbative expansion of the renormalized chemical
potential and the bulk modulus. The red solid line corresponds to the function $f_{1}(\theta)$
and the blue dashed line to the function $f_{2}(\theta)$. Both functions are symmetric around $\theta=\pi/2$. 
As we can see $f_{2}(\theta)$ is negative for all tilt angles. 
Therefore we can always find a critical interaction for which the bulk modulus becomes negative in second order perturbation theory.
}
\label{fig:functions_f}
\end{figure}

Given the renormalized Fermi surface $k_F ( \alpha )$, 
we may determine the renormalized
chemical potential $\mu = \gamma^2 E_{F0}$.
According to Luttingers theorem~\cite{Luttinger60}
the volume enclosed by the Fermi surface gives the 
particle density $n$. From the requirement that
the particle density must not change when the interaction is turned on
we obtain
\begin{align}
n=
\frac{1}{4\pi^2}\int\limits_{0}^{2\pi}   d \alpha 
 \int\limits_{0}^{k_{F}(\alpha)} d k k   = \frac{ k_{F0}^2}{4\pi}.
\end{align}
Substituting the self-consistency equation (\ref{eq:kfdef}) for $k_F ( \alpha )$
we obtain 
\begin{align}
 \mu
  & =   E_{F0}
 + \frac{2}{\pi} \int\limits_{0}^{\pi/2} d \alpha
  \Sigma( {k}_{F} (\alpha), 0 ).
\end{align}
Since we do not have analytic expressions for the Fermi momentum we evaluate $\mu$ numerically up to $u^2$. 
We find
 \begin{eqnarray}
 \frac{\mu}{E_{F0}} = 1 +f_{1}( \theta )u + f_{2}( \theta ) u^2 + {\cal{O}} ( u^3 ),
 \label{eq:mu}
 \end{eqnarray}
where the functions $f_{1}( \theta )$ and $f_{2}( \theta )$ are shown in Fig.~\ref{fig:functions_f}.
As we can see $f_{1} (\theta)$ is positive for small $\theta$ which is in the regime where the interaction is mostly repulsive.
In contrast, $f_{2} (\theta)$ is negative for all tilt angles, 
hence the second order contribution has a destabilizing effect for all $\theta$.

From the renormalized chemical potential we may determine the
bulk modulus
\begin{align}
 K=n^2 \left. \frac{\partial \mu}{\partial n} \right|_{V,T},
\end{align}
where the density can be expressed in terms of the
bare Fermi momentum as $n=k_{F0}^2/(4\pi)$.
We obtain
 \begin{equation}
 \frac{K}{  E_{F0}n } = 1 +\frac{3}{2}f_{1} ( \theta ) u + 2f_{2} ( \theta ) u^2 + {\cal{O}} ( u^3 )
 \label{eq:bulk2}
 \end{equation}
with the same functions $f_{1} ( \theta )$ and $f_{2} ( \theta )$ as in Eq.~\eqref{eq:mu}, see Fig.~\ref{fig:functions_f}.
Since $f_{2} ( \theta )$ is negative for all tilt angles we can always find a critical interaction
in second order perturbation theory for which the bulk modulus becomes negative -- even if the interaction is purely repulsive.
When the bulk modulus vanishes, the normal state of the system becomes unstable.
Using our second order result \eqref{eq:bulk2} to estimate the
instability regime, we obtain the critical tilt angle $\theta_c(u)$ for vanishing $K$.
The result is plotted in Fig.~\ref{fig:beta_1_of_u}.
For $\theta \lesssim \arcsin {(1/\sqrt{3})}$ our result for the instability line agrees quite well 
with the result of the random phase approximation for a density-wave instability derived by 
Yamaguchi \textit{et al.}~\cite{Yam10}.
The deviation between our critical interaction parameter and the critical interaction derived by Yamaguchi \textit{et al.}~\cite{Yam10} 
depends on the tilt angle, but is not larger than about $15\%$ in the regime $0 \le \theta \lesssim \arcsin {(1/\sqrt{3})}$.
Note that this instability and the breakdown of second order perturbation theory occur at a much smaller interaction strength than the instability to a stripe phase
predicted by Parish \textit{et al.}~\cite{Parish12}. 
Whether our result converges against the critical interaction derived by Parish \textit{et al.}~\cite{Parish12}
when the interaction is treated non-perturbatively is beyond the scope of this work.

We have also  calculated
the renormalized Fermi velocity, the quasi-particle residue and the
single-particle spectral function of our system. However,
in the regime where perturbation theory is valid the corrections are small
so that we do not present them here.

\section{Summary and Conclusion}
\label{sec:conclusion}

In this work we have calculated  the renormalized Fermi surface, the chemical potential, and
the bulk modulus in the normal state of two-dimensional dipolar fermions 
up to second order in perturbation theory.
We have pointed out that for idealized point dipoles  the
second-order self-energy
is ultraviolet divergent.
To regularize this divergence for electric dipoles, we have replaced the 
idealized point dipoles by 
physical dipoles with a finite length $\ell$.
Our regularization procedure takes all orders in the multipole expansion
into account and identifies the ultraviolet cutoff in momentum space
with the inverse dipole length.
The relevant regime where the dipole length $\ell$ is large compared 
with the spatial extend $a$ of the wave function in transverse direction, might at some point be 
experimentally accessible using dipolar Rydberg atoms or molecules.

While to first order in the interaction the renormalized Fermi surface is always
distorted along  the direction defined by the projection of the dipoles
onto the plane  of the system,
the second order correction tends to weaken this effect and, for a certain regime
of tilt angles $\theta$ and dimensionless interaction strengths $u = m d^2 k_{F0}$,
can even change the direction of distortion.
From our result for the renormalized chemical potential we have estimated
the bulk modulus and the parameter regime where the normal Fermi liquid is stable.
The qualitative behavior of our results does not depend on the regularization strategy.

Finally, let us emphasize that we have focussed on electric dipoles because in this case
the physical dipoles can be realized as two opposite charges which are
separated by a finite distance $\ell$. 
Obviously, our regularization is not valid for magnetic dipoles, however a similar procedure using
loop currents seems to be possible but is technically more cumbersome.

\section*{Acknowledgments}
We thank the DFG for financial support via FOR 723. 
We also thank Lorenz Bartosch for useful discussions and 
inspiring ideas.

\bibliographystyle{apsrev4-1}
\bibliography{paper_bib}

\end{document}